\documentclass{article}
\usepackage{graphics}
 \usepackage{graphicx}

 \usepackage{epsfig}

\usepackage{amssymb}
\usepackage{amsmath}
 \usepackage{amsthm}

\begin{document}

\title{Phonon-assisted tunneling through a double quantum dot system}

\author{M.˜ Bagheri Tagani,
        H.˜ Rahimpour Soleimani,\\
        \small{Department of physics, University of Guilan, P.O.Box 41335-1914, Rasht, Iran}}

\maketitle

\begin{abstract}
Electron transport  through a double quantum dot system is
studied with taking into account electron-phonon interaction. The
Keldysh nonequilibrium Green function formalism is used to
compute the current and transmission coefficient of the system.
The influence of the electron-phonon interaction, interdot
tunneling, and temperature on the density of states and current
is analyzed. Results show that although the electron-phonon
interaction results in the appearance of side peaks in the
conductance at low temperatures, they are disappeared in high
temperatures.
\end{abstract}

\section{Introduction}
\label{Introduction} The study of transport through systems
fabricated from two quantum dots has attracted a lot of attention
during last decades because of their molecular-like
behavior~\cite{van der
Wiel,Hanson1,Hayashi,Kieblich,Chi,Koppens,Molitor,Lim}.
Experimentally, quantum dots (QDs) can be in a serial or parallel
configuration~\cite{van der Wiel,Ono}. Discreteness of energy
levels, charging effects, and interdot tunneling in double
quantum dot (DQD) systems result in novel and interesting
phenomena such as: negative differential conductance
(NDC)~\cite{Chi,Weymann}, zero bias anomaly~\cite{Aguado,Chi1},
current rectification~\cite{Ono}, Pauli spin
blockade~\cite{Johnson}, ratchet effect~\cite{Moldoveanu}, etc.
Furthermore, a DQD system can be used to measure the lifetime of a
singlet-triplet state\cite{Prance}. DQD systems are also promising
candidates for spintronic applications~\cite{Awschalom} and
quantum computing~\cite{Loss,Hanson}.

\par Coupling between the electronic and vibrational degrees of
freedom can significantly affect the performance of the systems
fabricated from QDs. The electron-phonon interaction (EPI) in
nanostructures has been extensively studied both experimentally
and
theoretically~\cite{Park,LeRoy,Yang1,Wang1,Choi1,Dong,Galperin1,Lundin,Chen2,Liu1,Ueda}.
Different models are used to investigate the influence of the EPI
on the transport characteristics of the system such as: the rate
equation approach~\cite{Dong, Siddiqui}, the kinetic equation
method~\cite{McCarthy}, the Keldysh nonequilibrium Green function
formalism~\cite{Galperin1,Lundin,Chen2}, etc.
 Most work has been done on the single QD systems. However,
the EPI in the DQD systems is an important and interesting
subject needing more consideration.

\par In this work, we study the electron transport through a DQD
system using the Keldysh nonequilibrium Green function
formalism~\cite{Huag}. The issue has been recently analyzed by
means of a rate equation approach~\cite{Tagani}.  It is assumed
that the electrons of each dot interact with the vibrational
degrees of freedom.  The influence of the temperature, interdot
tunneling, and EPI strength on the spectral function of each dot
and transmission coefficient is examined. In the next section,
the model used to describe the system is presented. Sec. 3 is
devoted to the numerical results and in the end, some sentences
are given as a summary.

 \section{Model and formalism}
 \label{Model}
We consider two single level quantum dots attached to the
metallic electrodes and assume that the electrons of each dot
interact with a local vibrational mode. The Hamiltonian
describing the whole system is given as ($\hbar=1$)
\begin{align}\label{Eq.1}
H=& \sum_{\alpha k}\varepsilon_{\alpha k} c^{\dag}_{\alpha
k}c_{\alpha k}+\sum_{i=L,R} \varepsilon_{i}
d^{\dag}_{i}d_{i}+t[d^{\dag}_{L}d_{R}+H.C]+\\\nonumber
&\sum_{i}[\omega_i a^{\dag}_{i}a_{i}+\lambda_{i}
[a^{\dag}_{i}+a_{i}]n_{i}]+\sum_{\alpha k i}[V_{\alpha k
i}c^{\dag}_{\alpha k} d_{i}+H.C]
\end{align}
where $c^{\dag}_{\alpha k}$ ($c_{\alpha k}$) creates (destroys)
an electron with wave vector $k$, energy $\varepsilon_{\alpha
k}$, in lead $\alpha$ whereas, $d^{\dag}_{i}$ ($d_{i}$) is the
creation (annihilation) operator in the $i$th dot. The third term
describes the interdot tunneling and $t$ denotes the interdot
tunneling strength. $\omega_i$ is the phonon energy in the dot
$i$, while $\lambda_i$ stands for EPI strength in the $i$th dot.
We assume that each dot can have up to one electron. With respect
to the fact that the energy levels of the QD are controlled using
a gate voltage and, on the other hand, the on-site Coulomb
repulsion is order of a few $meV$, the assumption is reasonable.
Assuming weak lead-dot coupling, the EPI can be eliminated using
nonperturbative canonical transformation~\cite{Mahan},
$\tilde{H}=e^{S}He^{-S}$, where
$S=\sum_{i}\lambda_i/\omega_i[a^{\dag}_{i}-a_{i}]n_{i}$.
Therefore, Eq.\eqref{Eq.1} becomes
\begin{align}\label{Eq.2}
\tilde{H}=& \sum_{\alpha k}\varepsilon_{\alpha k} c^{\dag}_{\alpha
k}c_{\alpha k}+\sum_{i=L,R} \tilde{\varepsilon}_{i}
d^{\dag}_{i}d_{i}+t[X^{\dag}_{L}X_{R}d^{\dag}_{L}d_{R}+H.C]+\\\nonumber
&\sum_{i}\omega_i a^{\dag}_{i}a_{i}+\sum_{\alpha k i}[V_{\alpha k
i}X_{i}c^{\dag}_{\alpha k} d_{i}+H.C]
\end{align}
where
$\tilde{\varepsilon}_{i}=\varepsilon_{i}-\lambda^2_{i}/\omega_i$
is the renormalized energy level of the $i$th dot due to the
polaronic shift and,
$X_{i}=e^{-\frac{\lambda_i}{\omega_{i}}[a^{\dag}_{i}-a_{i}]}$ is
the phonon shift generator operator. Since the dot-lead coupling
strength is weaker than the EPI strength, i.e., $V_{\alpha
ki}<<\lambda_i$, $X_i$ is replaced with its expectation value,
i.e., $X_i=<X_i>=e^{-g_i(2N_{ph,i}+1)}$, where
$g_i=(\lambda_i/\omega_i)^2$, and $N_{ph,i}$ denotes the phonon
population expressed as $N_{ph,i}=(exp((\beta\omega_i)-1)^{-1}$
with $\beta=1/k_B T$. In the following, it is assumed that both
QDs have the same shape and size so that the phonon energy and
EPI strength are independent of the QDs indexes and are shown by
$\omega_0$, and $\lambda$, respectively.
\par In order to compute the current, the Keldysh nonequilibrium
Green function formalism and wide band approximation are used so
that the current can be expressed as~\cite{Meir}
\begin{equation}\label{Eq.3}
  I=\frac{e}{2h}\int d\omega \{Tr[(f_{L}(\omega)\Gamma^{L}-f_{R}(\omega)\Gamma^{R})A(\omega)]+Tr[(\Gamma^{L}-\Gamma^{R})(iG^{<}(\omega))] \}
\end{equation}
where
$f_{\alpha}(\omega)(1+exp((\omega-\mu_{\alpha})\beta))^{-1}$ is
the Fermi distribution function of the lead $\alpha$, and
$\mu_{\alpha}$ denotes the chemical potential of the lead.
$A(\omega)$ and $G^{<}(\omega)$ are the Fourier transformations
of the spectral function  and lesser Green function,
respectively. The matrix $\Gamma^{\alpha}$ describes the
tunneling coupling between the $\alpha$th dot and the leads given
as
\begin{equation}\label{Eq.4}
  \Gamma^{L}=\begin{pmatrix}
    {\Gamma_0} & {\sqrt{\alpha} \Gamma_0} \\
    {\sqrt{\alpha}\Gamma_0} & {\alpha\Gamma_0} \
  \end{pmatrix}
  \quad \Gamma^{R}=\begin{pmatrix}
    {\alpha\beta\Gamma_0} & {\beta\sqrt{\alpha} \Gamma_0} \\
    {\beta\sqrt{\alpha}\Gamma_0} & {\beta\Gamma_0} \
  \end{pmatrix}
\end{equation}
where $\Gamma_0$ is a constant, $\alpha$ describes the difference
in the coupling of the electrodes to different QDs, and $\beta$
stands for asymmetry in the coupling of the QDs to the left and
right leads. The spectral function is computed as
\begin{equation}\label{Eq.5}
  A(\omega)=i(G^{>}(\omega)-G^{<}(\omega))=i(G^{r}(\omega)-G^{a}(\omega))
\end{equation}
where $G^{r(a)}(\omega)$ is the retarded (advanced)  Green
function of the whole system. To compute $G^{<(>)}(\omega)$ we
follow the procedure introduced in Ref.~\cite{Chen2}. Because the
phonon shift generator operator is replaced with its expectation
value, the Green functions can be separated to the electron and
phonon parts as
\begin{subequations}
\begin{align}\label{Eq.6}
  G^{>}_{ij}(t)&=-i<d_i(t)d^{\dag}_{j}(0)>=\tilde{G}^{>}_{ij}(t)e^{-\Phi(t)}\\
  G^{<}_{ij}(t)&=i<d^{\dag}_{j}(0)d_i(t)>=\tilde{G}^{<}_{ij}(t)e^{-\Phi(-t)}
\end{align}
\end{subequations}
where $i(j)=1,2$ stands for the matrix elements,
$\tilde{G}^{<(>)}(t)$ is the dressed lesser (greater) Green
function resulting from the electronic part of Eq.\eqref{Eq.2},
and $e^{-\Phi(t)}$ comes from the phonon part,
$<X(t)X^{\dag}(0)>$, given as~\cite{Mahan}
\begin{equation}\label{Eq.7}
  e^{-\Phi(t)}=g[N_{ph}(1-e^{i\omega_0 t})+(N_{ph}+1)(1-e^{-i\omega_0 t})]
\end{equation}
\par Using the identity
$e^{-\Phi(t)}=\sum_{n}L_{n}e^{-in\omega_0t}$, $G^{<(>)}(\omega)$
can be expressed as
\begin{subequations}\label{Eq.8}
  \begin{align}
  G^{>}(\omega)&=\sum_{n}L_{n}\tilde{G}^{>}(\omega-n\omega_0)\\
  G^{<}(\omega)&=\sum_{n}L_{n}\tilde{G}^{<}(\omega+n\omega_0)
  \end{align}
\end{subequations}
where $L_n$ is a function of temperature, T, EPI strength, and
phonon population given as~\cite{Chen2}
\begin{align}\label{Eq.9}
 L_n&=\frac{e^{-g}g^{n}}{n!} \quad for \quad n \succeq 0 \quad if
 \quad T=0\\\nonumber
 L_n&=e^{-g(2N_{ph}+1)}e^{\frac{n\omega_0\beta}{2}}I_{n}(2g\sqrt{N_{ph}(N_{ph}+1)})\quad
 if \quad T \neq 0
\end{align}
where $I_n$ is the $n$th Bessel function of complex argument.
$\tilde{G}^{<}(\omega)$ can be computed using the Keldysh
equation and Langreth analytical continuation as
$\tilde{G}^{<}(\omega)=\tilde{G}^{r}(\omega)\tilde{\Sigma}^{<}(\omega)\tilde{G}^{a}(\omega)$,
where
$\tilde{\Sigma}^{<}(\omega)=i[\tilde{\Gamma}_{L}f_{L}(\omega)+\tilde{\Gamma}_{R}f_{R}(\omega)]$,
and $\tilde{\Gamma}_{\alpha}=\Gamma_{\alpha}<X>^2$. Indeed,
contribution to $\Sigma^{<}$ from electron-phonon interaction is
disregarded due to weak electron-phonon interaction. In order to
compute the dressed retarded Green function, the equation of
motion technique is used. It is straightforward to show that the
matrix elements of the Green function satisfy the following
equation
\begin{subequations}\label{Eq.10}
\begin{align}
  [\omega-\tilde{\varepsilon}_{i}-\tilde{\Sigma}_{ii}]\tilde{G}^{r}_{ii}(\omega)&=1+(t+\tilde{\Sigma}_{i\bar{i}})\tilde{G}^{r}_{\bar{i}i}(\omega)\\
  [\omega-\tilde{\varepsilon}_{i}-\tilde{\Sigma}_{ii}]\tilde{G}^{r}_{i\bar{i}}(\omega)&=(t+\tilde{\Sigma}_{i\bar{i}})\tilde{G}^{r}_{\bar{i}\bar{i}}(\omega)
\end{align}
\end{subequations}
where $\bar{i}=2$ if $i=1$,  $\bar{i}=1$ if $i=2$, and
$\tilde{\Sigma}=-i/2[\tilde{\Gamma}_{L}+\tilde{\Gamma}_{R}]$.
Indeed, the real part of the self energy causing slightly shift
in the QD energy levels is ignored (wide band approximation).
With respect to Eq.\eqref{Eq.10}, the retarded Green function is
obtained as
\begin{equation}\label{Eq.11}
  \tilde{G}^{r}(\omega)=\frac{1}{P(\omega)}\begin{pmatrix}
    \omega-\tilde{\varepsilon}_{2}-\tilde{\Sigma}_{22} & t+\tilde{\Sigma}_{12} \\
    t+\tilde{\Sigma}_{21} & \omega-\tilde{\varepsilon}_{1}-\tilde{\Sigma}_{11} \
  \end{pmatrix}
\end{equation}
where
\begin{equation}\label{Eq.12}
  P(\omega)=(\omega-\tilde{\varepsilon}_{1}-\tilde{\Sigma}_{11})(\omega-\tilde{\varepsilon}_{2}-\tilde{\Sigma}_{22})-
  (t+\tilde{\Sigma}_{12})(t+\tilde{\Sigma}_{21})
\end{equation}
The Green function of the whole system can be written
as~\cite{Rudzinsky}
\begin{align}\label{new}
  G^{r}_{ij}(t)&=-i\Theta(t)<\{d_{i}(t),d^{\dag}_{j}(0)\}>=\tilde{G}^{r}_{ij}(t)<X(t)X^{\dag}(0)>\\\nonumber
  &-\Theta(t)\tilde{G}^{<}_{ij}(t)[<X^{\dag}(0)X(t)>-<X(t)X^{\dag}(0)>]
\end{align}
where after Fourier transformation, it becomes
\begin{equation}\label{Eq.13}
  G^{r}(\omega)=\sum_{n}L_{n}[\tilde{G}^{r}(\omega-n\omega_0)-\frac{1}{2}\tilde{G}^{<}(\omega+n\omega_0)+\frac{1}{2}\tilde{G}^{<}(\omega-n\omega_0)]
\end{equation}

\par For simulation purposes, $\omega_0$ is used as energy unit~\cite{explain1},
and it is assumed that the energy levels of the QDs are
degenerate so that the density of states of the QDs are identical
shown by $A(\omega)$.

\section{Results and discussions }
\label{Numerical results} Fig. 1 shows the density of states as a
function of energy for different $\lambda$s and $t$s. In fig. 1a
that $t=\Gamma_0$, there is a delta-like peak in the bonding
state ($\tilde{\varepsilon_i}-t$) and a Lorentzian peak in the
antibonding state ($\tilde{\varepsilon_i}+t$). For weak
$\lambda$s, the below relation is obtained for $A(\omega)$ at the
bonding and antibonding states
\begin{equation}\label{Eq.14}
  A(\tilde{\varepsilon}_i\pm t)=4\tilde{\Gamma_0}\frac{8t^2(1\pm
  \sqrt{\alpha})^2+\tilde{\Gamma_0}^{2}(1+\alpha)(1-\alpha)^2}{\tilde{\Gamma_0}^4(1-\alpha)^4+16(1\pm \sqrt{\alpha})^{4}t^2\tilde{\Gamma_0}^2}
\end{equation}
For strong $\lambda$s, the bonding and antibonding states are
farther from each other due to the red shift. Furthermore,
satellite peaks are well observed in strong EPI coming from the
phonon emission by electrons (at negative energies) and the
phonon emission by holes (at positive energies). Note that the
probability of the phonon absorption is zero in zero temperature
because of $N_{ph}=0$. As one expects, the spacing between the
satellite peaks is equal to the phonon energy. The influence of
the interdot tunneling strength on $A(\omega)$ is analyzed in
fig. 1b. At $t/\Gamma_0<<1$, the bonding and antibonding states
are merged. With increase of $t$, these states are well
separated. It is interesting to note that although the density of
states are independent of the chemical potential of the leads in
elastic transport, it is dependent on the position of the
chemical potential of the leads in the presence of the EPI
according to Eq.\eqref{Eq.5}. However, the dependence does not
affect significantly on $A(\omega)$.
\par The influence of the temperature on $A(\omega)$ is plotted in
fig. 2. With increase of the temperature, the height of the peaks
is increased whereas, their width is reduced. Such behavior was
previously reported about a single level QD~\cite{Chen2}. The
dressing effect is increased by increase of $T$, so that
$\tilde{\Gamma}$ becomes smaller and, with respect to the fact
that the width of the spectral function is related to the
broadening, increase of the temperature results in the reduction
of the width of the peaks. In addition, the probability of the
phonon absorption is increased by increasing $T$ and, as a
result, the satellite peaks in $A(\omega)$ result from both the
phonon emission and the phonon absorption. Unlike the elastic
transport, the spectral function depends on the Fermi
distribution of the leads in inelastic transport so that height
of the peaks is increased by increase of the temperature owing to
spreading the Fermi function by increase of $T$.
\par Transmission coefficient,
$T(\omega)=Tr(G^{a}{\Gamma}^{R}G^{r}{\Gamma}^{L})$, is shown in
fig. 3 in zero temperature. There are two main peaks located in
the bonding and antibonding states and a antiresonance-like
behavior due to the destructive interference between different
pathways through the system.  The effect was before announced for
a coupled QD system~\cite{Guevara}. With increase of the EPI
strength, the position of the bonding and antibonding states is
shifted toward left. Furthermore, the secondary peaks are well
observed resulting from the existence of the new channels for the
electron transport because of the electron-phonon coupling. The
probability of the electron transport through these channels is
lesser than the probability of the electron transport through the
main channels. The effect of the dot-lead coupling on $T(\omega)$
is studied in fig. 3b. Such analysis was before done without
considering the EPI~\cite{Guevara}. In serial configuration
($\alpha=0$), the transmission coefficient exhibits two
Lorentzian peaks centered at the bonding and antibonding states,
respectively. In addition, the satellite peaks due to the
phonon-assisted tunneling are clearly observed. With increase of
$\alpha$, the antiresonance-like behavior is also observed
because of the destructive interference.
\par Fig. 4 shows the current-voltage characteristic of the system
for different temperatures. In zero temperature, the current
exhibits step-like behavior so that there is a step when the
bonding or antibonding state is located inside the bias window.
More steps are observed in the presence of the EPI because of the
phonon-assisted tunneling. The side peaks are well seen in the
conductance spectrum. With increase of the temperature, the
step-like behavior is vanished. Indeed, the electrons gain more
energy to tunnel through the system owing to the thermal
excitation. On the other hand, the effects coming from the EPI
are not observed in the high temperature so that there is only
one peak in the conductance.

\section{Summary}
\label{conclusion} The Keldysh nonequilibrium Green function
formalism is used to study the electron transport through a
double single level quantum dot system in the presence of the
electron-phonon interaction. The behavior of the system is
analyzed in both zero and nonzero temperatures using the
nonperturbative canonical transformation. The current voltage
characteristic exhibits the step-like behavior in low
temperatures  because of the EPI whereas, such behavior is
disappeared at high temperatures. The influence of the
temperature, EPI strength, and interdot tunneling strength on the
transmission coefficient and density of states is also examined.

\bibliographystyle{model1a-num-names}
\bibliography{<your-bib-database>}

\newpage
\textbf{Figure captions}

\par Figure 1: Density of states as a function of energy in zero temperature.
Parameters are $\varepsilon_i=0$, $\mu_\alpha=0$, $\alpha=0.4$,
$\beta=1$, and $\Gamma_0=0.2$.

\par Figure 2: $A(\omega)$ versus energy for different
temperatures. $\lambda=0.5$, and $t=0.2$. Other parameters are
the same as fig. 1.

\par Figure 3: Transmission coefficient as a function of energy.
Parameters are the same as fig. 1. $\lambda=0.5$ in fig. b.

\par Figure 4: The current versus voltage for $\lambda=0.5$ (solid
line) and $\lambda=0$ (dashed line) at $kT=0$ (gray) and $kT=1$.
Inset shows the conductance. Parameters are $\varepsilon_i=3$,
$t=2\Gamma_0=0.4$.

\newpage

\begin{figure}[htb]
\begin{center}
\includegraphics{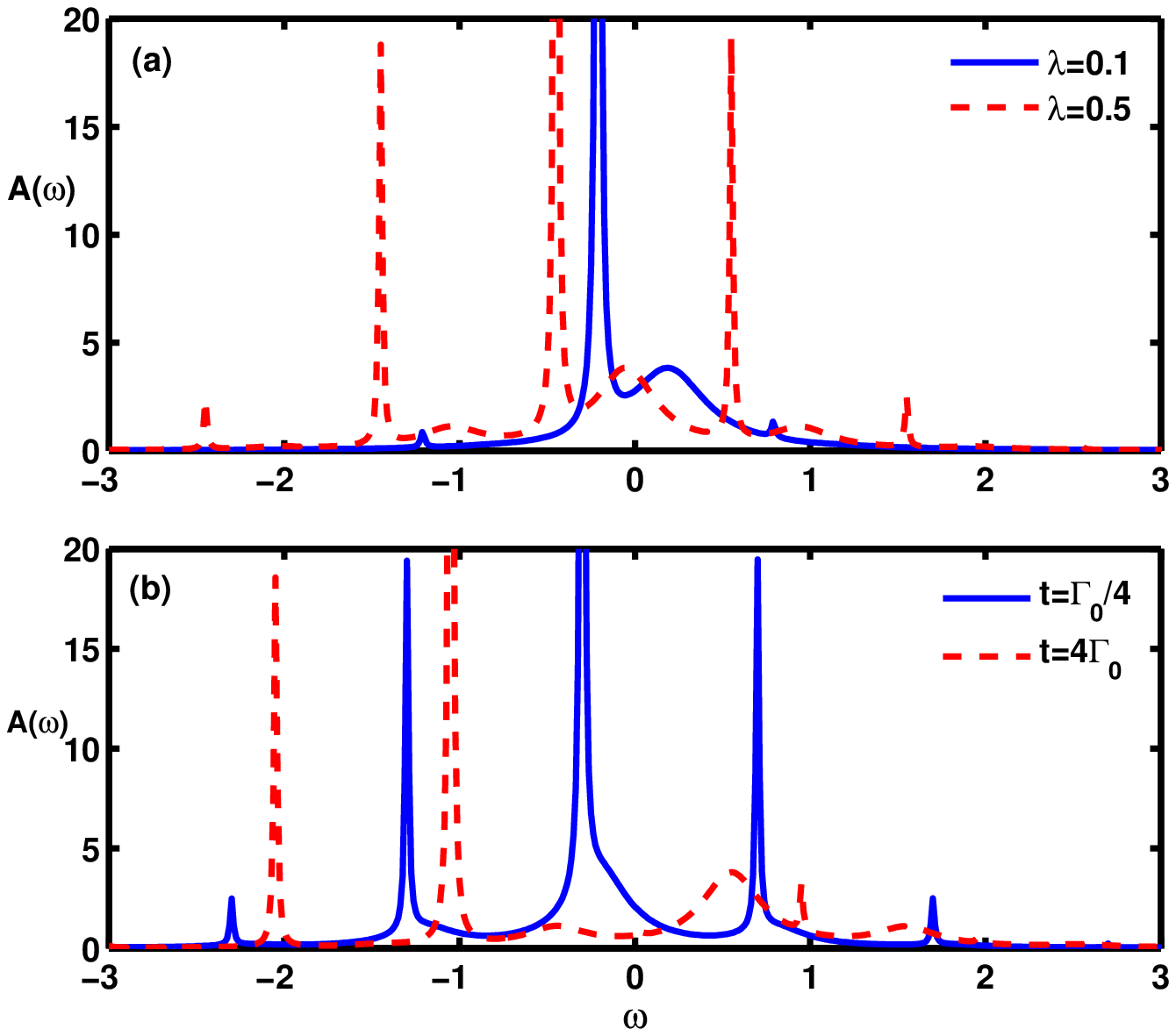}\nonumber
\caption{}\label{fig:1}       
\end{center}
\end{figure}

\begin{figure}[htb]
\begin{center}
\includegraphics{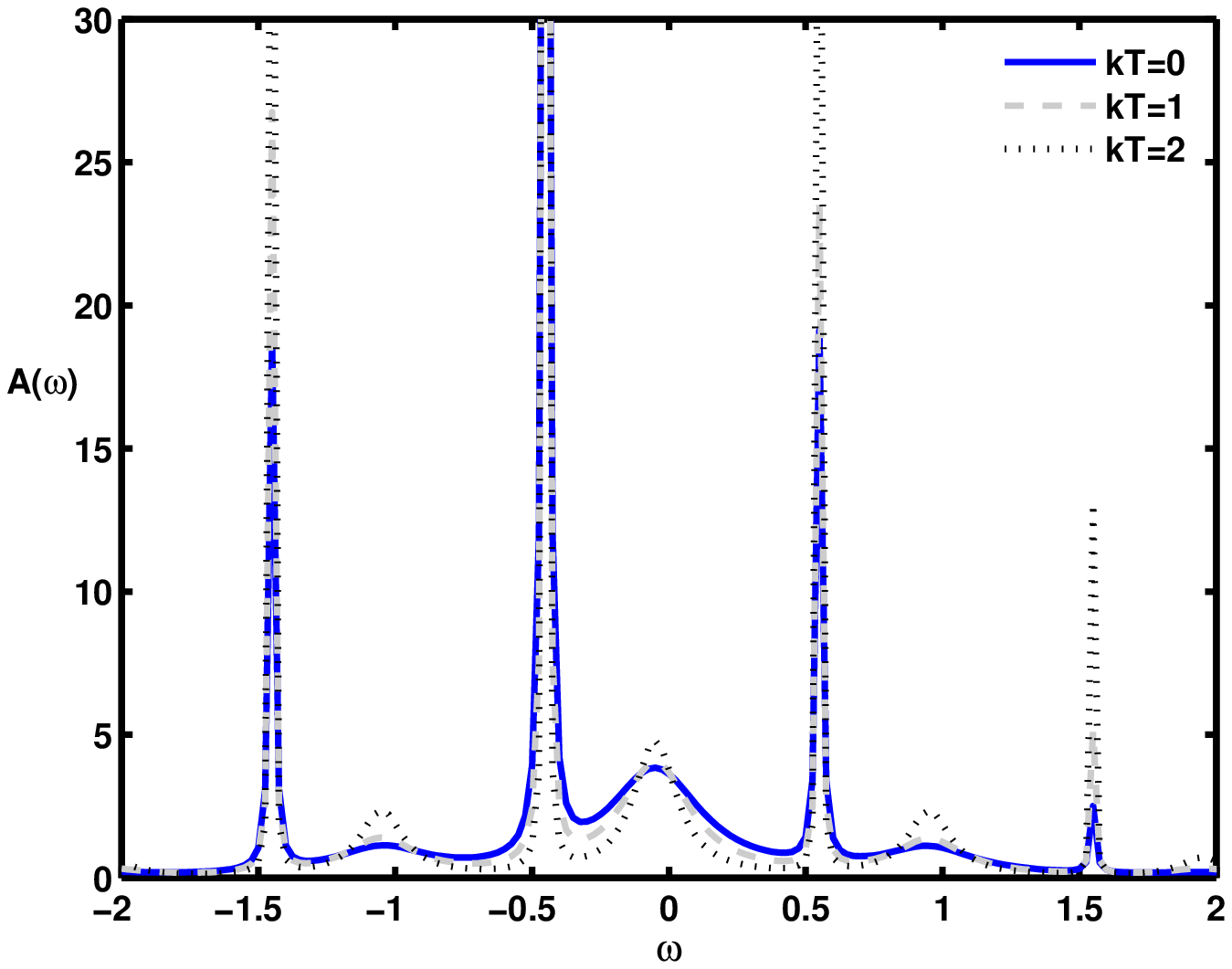}\nonumber
\caption{}\label{fig:2}       
\end{center}
\end{figure}

\begin{figure}[htb]
\begin{center}
\includegraphics{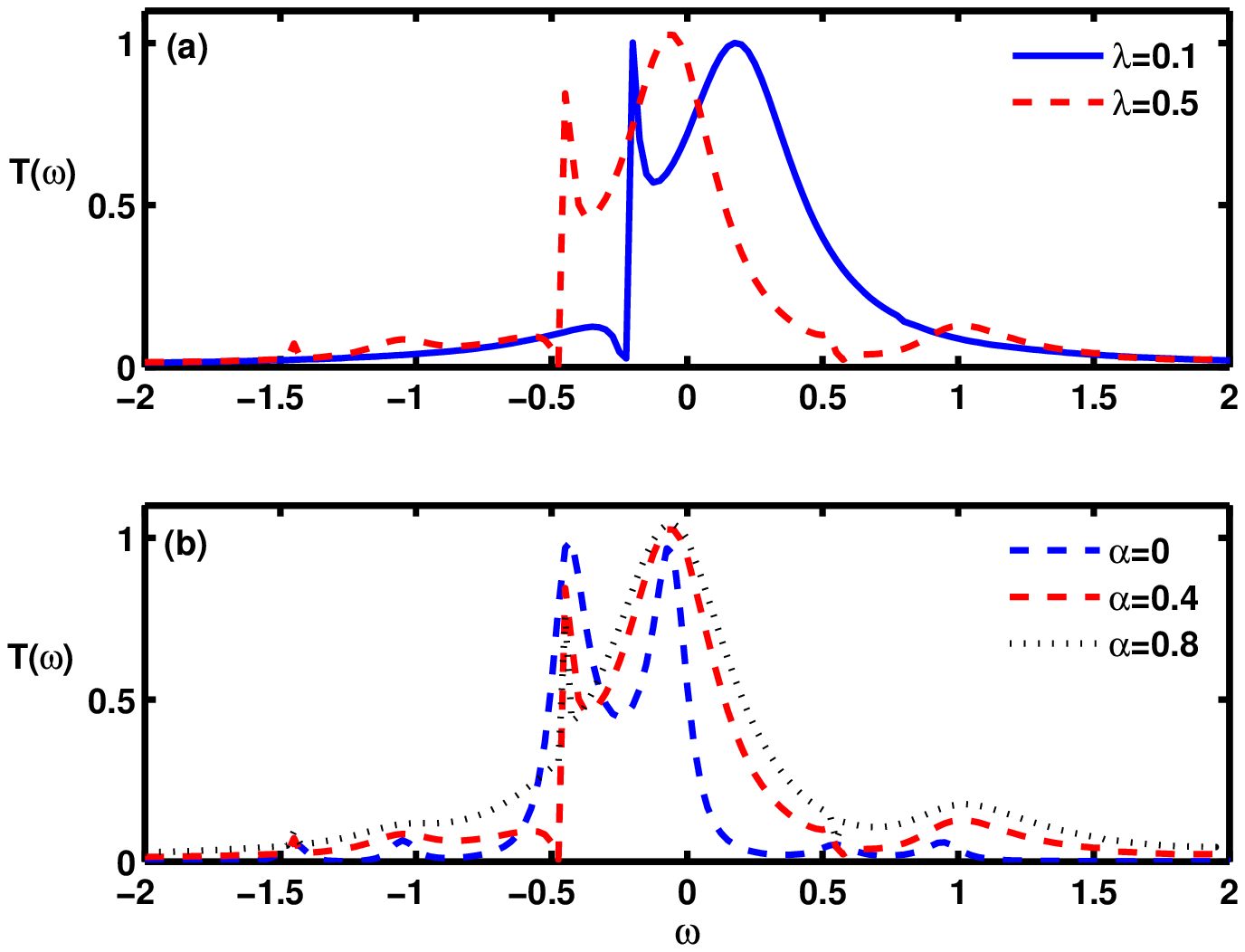}\nonumber
\caption{}\label{fig:3}       
\end{center}
\end{figure}

\begin{figure}[htb]
\begin{center}
\includegraphics{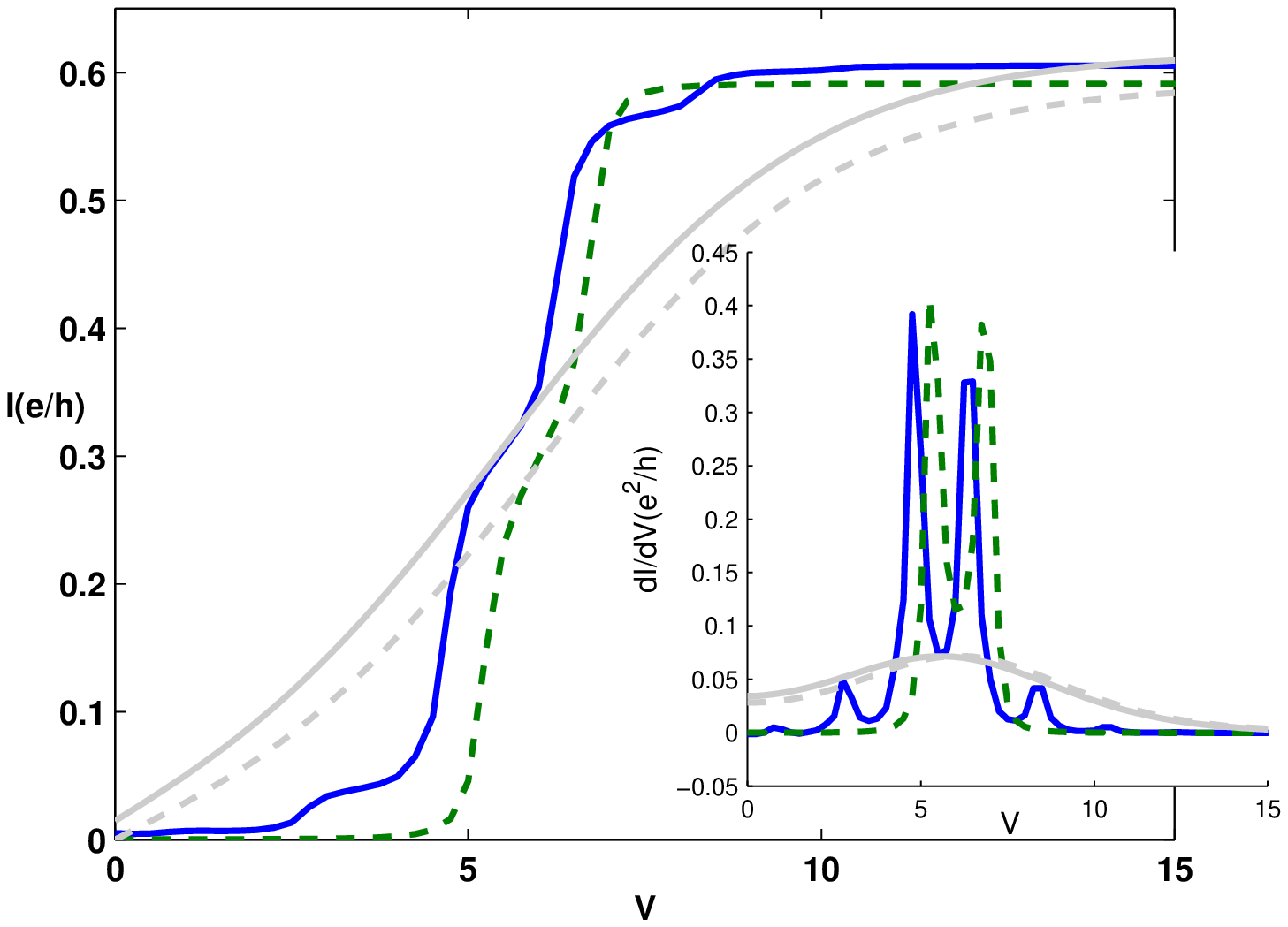}\nonumber
\caption{}\label{fig:4}       
\end{center}
\end{figure}

\end{document}